\documentclass[fleqn,usenatbib]{mnras}

\usepackage{newtxtext,newtxmath}

\usepackage[T1]{fontenc}

\DeclareRobustCommand{\VAN}[3]{#2}
\let\VANthebibliography\thebibliography
\def\thebibliography{\DeclareRobustCommand{\VAN}[3]{##3}\VANthebibliography}

\usepackage{graphicx}	
\usepackage{amsmath}	
\usepackage{xcolor}

\title[The impact of incorrect dissociation energies]{The impact of incorrect dissociation energies on inferred photospheric abundances}

\author[Aquilina et al.]{
Sarah E. Aquilina$^{1}$,
Andrew R. Casey$^{1,2}$,
Adam J. Wheeler$^{3}$\\
$^{1}$School of Physics and Astronomy, Monash University, VIC 3800, Australia\\
$^{2}$ARC Centre of Excellence for All Sky Astrophysics in Three Dimensions (ASTRO-3D)\\
$^{3}$Department of Astronomy, Ohio State University, McPherson Laboratory, 140 West 18th Avenue, Columbus, OH USA\\
}

\date{Accepted XXX. Received YYY; in original form ZZZ}

\pubyear{2023}

\begin{document}
\label{firstpage}
\pagerange{\pageref{firstpage}--\pageref{lastpage}}
\maketitle

\begin{abstract}
Spectral synthesis codes are essential for inferring stellar parameters and detailed chemical abundances. These codes require many physical inputs to predict an emergent spectrum. Developers adopt the best measurements of those inputs at the time they release their code, but those measurements usually improve over time faster than the software is updated. In general, the impact of using incorrect or uncertain dissociation energies are largely unknown.
Here we evaluate how incorrect dissociation energies impact abundances measured from C$_2$, CN, CH, TiO, and MgO features.
For each molecule we synthesised optical spectra of FGKM-type main-sequence and giant stars using the literature dissociation energy, and an incorrect (perturbed) dissociation energy. 
We find that the uncertainties in the dissociation energies adopted by spectral synthesis codes for CN, CH, TiO, and MgO lead to negligible differences in flux or abundances. C$_2$ is the only diatomic molecule where the uncertainty of the inputted dissociation energy translates to a significant difference in flux, and carbon abundance differences of up to 0.2 dex. For Solar-like stars, the impact on carbon abundance is up to 0.09 dex. These large abundance differences demonstrate the importance of updating the inputs adopted by spectral synthesis codes, as well as a consensus on appropriate values between different codes.

\end{abstract}


\section{Introduction}
\label{Section: introduction}
Stellar spectra are necessary to understand production sites of elements \citep{jofre_gaia_2015}, the chemical enrichment of galaxies \citep{amarsi_non-lte_2016}, and even to understand the formation of exoplanets \citep{wang_detailed_2022}, among others. However, generating synthetic (model) spectra for comparison requires numerous uncertain physical inputs. While numerous comparisons exist between spectral synthesis codes \citep[e.g.,][]{ges}, no detailed review has evaluated how incorrect physical quantities or their uncertainties translate to systematic uncertainties in stellar parameters or elemental abundances.
\\\\
The collective impact of incorrect physical parameters in modelling stellar spectra accurately is highlighted by inconsistent [C/Fe] abundances in carbon-enhanced metal-poor (CEMP) stars from various stellar surveys \citep{arentsen_inconsistency_2022}. CEMP stars are very old stars that contain vital information about the early Universe. They also have strong carbon molecular features that are routinely used to estimate the photospheric carbon abundance. This is used to classify CEMP stars by the typical definition of $[\mathrm{C/Fe}] > +1$ (in some cases $[\mathrm{C/Fe}] > 0.7$) and $[\mathrm{Fe/H}] < -1$ \citep{beers_discovery_2005}. However, various surveys such as the Solan Digital Sky Survey (SDSS), the All-Wavelength Extended Growth Strip International Survey (AEGIS), the SkyMapper Survey and Large Sky Area Multi-Object Fibre Spectroscopic Telescope (LAMOST) all obtain different distributions of [C/Fe] abundances, leading to different fractions of CEMP stars for each individual stellar populations \citep{arentsen_inconsistency_2022}. These disagreements translate to contradicting inferences about the conditions of the early universe, and how star formation proceeded at high redshift. While \citet{arentsen_inconsistency_2022} is comparing two different pipelines that use different C$_2$ bands, such discrepancies might be reduced by a consistent use of updated physical inputs needed by spectral synthesis codes for predicting synthetic spectra, which we then compare to observations.
\\\\
One quantity required by spectral synthesis codes is the dissociation energy for each molecular species considered, a requirement for computing molecular equilibrium constants. The dissociation energy is the energy needed to break molecular bonds, increasing the number of particles. The dissociation energy is preferably determined experimentally, as there are specific methods for different molecules. For diatomic molecules, spectrometry provides the most accurate dissociation energies \citep{luo2007comprehensive}. This could be through band convergence, extrapolation to convergence limits, the long-wavelength limit of absorption continuum, predissociation limits and/or photodissociation \citep{luo2007comprehensive}. However, only band convergence and extrapolation provide values for the dissociation energy, while the other methods typically provide upper limits \citep{wilkinson_1963}.
\\\\
When experimental measurements are unavailable, there are multiple ways in which the dissociation energy can be theoretically determined: algebraic methods \citep{weiguo_accurate_2007}, Gaussian-1,2,3,4 (G1, G2, G3, G4) theory \citep{pople_gaussian1_1989} and coupled cluster \citep{raghavachari_augmented_1985} methods. Gaussian theory is generally preferred for stellar purposes because it can be applied to a larger range of molecules compared to coupled cluster methods and algebraic methods \citep{curtiss_gaussian-4_2007}. Coupled cluster calculations are prohibitively expensive for complex molecules and for all molecules could lead to large errors than second order Møller-Plesset perturbation theory \citep{bartlett_1974} in G1 and G2 theory. However, different coupled cluster calculations are sometimes implemented in Gaussian theory to account for core valence effects, relativistic effects, and atomic spin-orbit effects \citep{curtiss_gaussian-4_2007}. The extent to which these effects are included is dependent on the approximations of each Gaussian theory. 
\\\\
To overcome some of the limitations of these older theoretical methods, recent dissociation energy values have been determined using a combination of multi-reference configuration interaction calculations and four-wave mixing experiments \citep{visser_new_2019}. This implements an improved description of the ground state wave function for molecules such as C$_2$ and includes the highest electronic states. As a result, the dissociation energy values stated by \citet{visser_new_2019} have smaller systematic errors than the other theoretical methods mentioned.
\\\\
Despite recent dissociation energy values being more accurate and precise \citep[eg.][]{visser_new_2019,ruscic_active_2014}, spectral synthesis codes typically use collated tables assembled for the purpose, which may be significantly out of date. The choice of dissociation energy values will impact the molecular opacity and subsequently the atomic abundances derived from the opacity to a greater extent than partition functions \citep{costes_naulin_1994}. Updating the atomic and molecular data implemented in spectral synthesis codes will ensure they all use the same precise values, reducing systematic errors and reconciling discrepancies between theoretical results.
\\\\
The most recent compendium of dissociation energies for use in stellar spectral synthesis and atmosphere modeling can be found in \citet{barklem_partition_2016}. This data set considers dissociation energies from experiments of \citet{luo2007comprehensive}, theoretical values from G2 theory \citep[e.g.,][]{curtiss_gaussian2_1991} as well as values from an older compendium by \citet{herzberg1950molecular}. For each molecule, a dissociation energy was adopted from one of these sources. Some energies from \citet{luo2007comprehensive} were taken from \citet{herzberg1950molecular}, and agreed well with \citet{curtiss_gaussian2_1991}. However, \citet{luo2007comprehensive} provided error estimates on dissociation energies, which is critically important for this work, and therefore was chosen in most cases. These values are adopted by \texttt{Korg}, a new spectral synthesis code \citep{korg_2022}, which motivates our interest in checking their accuracy and impact on resultant spectra. Other synthesis codes use values originating from older data sources which have been augmented over time, though they are nearly universally less recent.
\\\\
The paper is structured as follows. In Section \ref{Section: method} we describe the inputs used to synthesise spectra with `fiducial' and `incorrect' (perturbed) dissociation energies. This includes details of how we evaluate the impact that the perturbed dissociation energy has on the spectrum. In Section \ref{Section: results} we quantify the discrepancies from perturbed dissociation energies as flux differences and abundance differences. These differences between the fiducial and perturbed spectra are discussed in detail in Section \ref{Section: discussion}.  

\section{Methods}

\label{Section: method}

In this work we investigated five molecules: C$_2$, CN, CH, TiO and MgO. These molecules are important in stellar astrophysics: they contribute significantly to the opacity, and many features can be readily observed in stellar spectra. These five molecules also have dissociation energies with uncertainties greater than 0.01~eV 
in \citet{barklem_partition_2016}, which is larger than the typical uncertainties included in that work. Dissociation energies and uncertainties adopted for each molecule are shown in Table~\ref{table:dissociation energies}. 
\\\\
We synthesised spectra from 3000\,\AA\ to 9000\,\AA\ using \texttt{Korg} \citep{korg_2022} and a line list from the Vienna Atomic Line Database \citep[][and references therein]{kupka_1999}. We computed spectra using a grid of MARCS model atmospheres \citep{gustafsson_grid_2008} with effective temperatures from 3000\,K to 8000\,K in 250\,K intervals, surface gravities from 0.0 to 5.5 in 0.5\,dex intervals, $[\mathrm{Fe/H}]$ from $-1.5$ to 0.5 in 0.5 intervals, [$\alpha$/H] from 0.0 to 0.4 in 0.1 intervals, [C/metals] = 0.0 and [N/metals] $= 0.0$. Some atmospheres had a diverging chemical equilibrium and therefore the corresponding spectrum was not produced.
\\\\
For each molecule under examination, we computed two spectra for each photosphere: one using the `fiducial' dissociation energy (from \citealp{barklem_partition_2016}), and one using an 'incorrect' (perturbed) dissociation energy: the fiducial values minus their uncertainty  \citep[see Table~\ref{table:dissociation energies} or][]{barklem_partition_2016}.
We assume that the impact of perturbed dissociation energies on the emergent spectrum (or the inferred abundance) would be symmetric, such that we do not synthesise spectra with a perturbed dissociation energy where the uncertainty is instead added.
\\\\
We did not independently analyse the impact of discrepant dissociation energies from various spectral synthesis codes (\texttt{MOOG} \citep{sneden_2012}, \texttt{SME} \citep{SME_1996}, and \texttt{TurboSpectrum} \citep{turbospectrum_2012}), as the dissociation energy values adopted for the five molecules in our investigation are all within one standard deviation of those in \citet{barklem_partition_2016} (except for MgO in \texttt{MOOG}). Perturbing the dissociation energies by their uncertainty allows us to quantify the largest impact that incorrect dissociation energies could have on inferred chemical abundances. Each of these spectral synthesis codes also have other inherent differences besides physical inputs such as continuum normalisation, treatment of line broadening, and interpolation of model atmospheres. These variations may affect the abundances that we infer \citep[eg.][]{jofre_gaia_2017} and skew the impact of incorrect dissociation energies that we observe.
\\
\begin{table}
\centering
\begin{tabular}{|c|c|c|} 
 \hline
  Molecule & Dissociation Energy (eV) & Uncertainty (eV) \\
 \hline
  C$_2$    & 6.371 & 0.160 \\
  CN       & 7.737 & 0.030 \\
  CH       & 3.4696 & 0.013 \\
  TiO      & 6.869  & 0.061 \\
  MgO      & 3.673 &  0.074 \\
 \hline
\end{tabular}

\caption[]{Dissociation energies and uncertainties from \citet{barklem_partition_2016} adopted for each molecule.}
\label{table:dissociation energies}
\end{table}

\begin{figure*}
	\includegraphics[width=\textwidth]{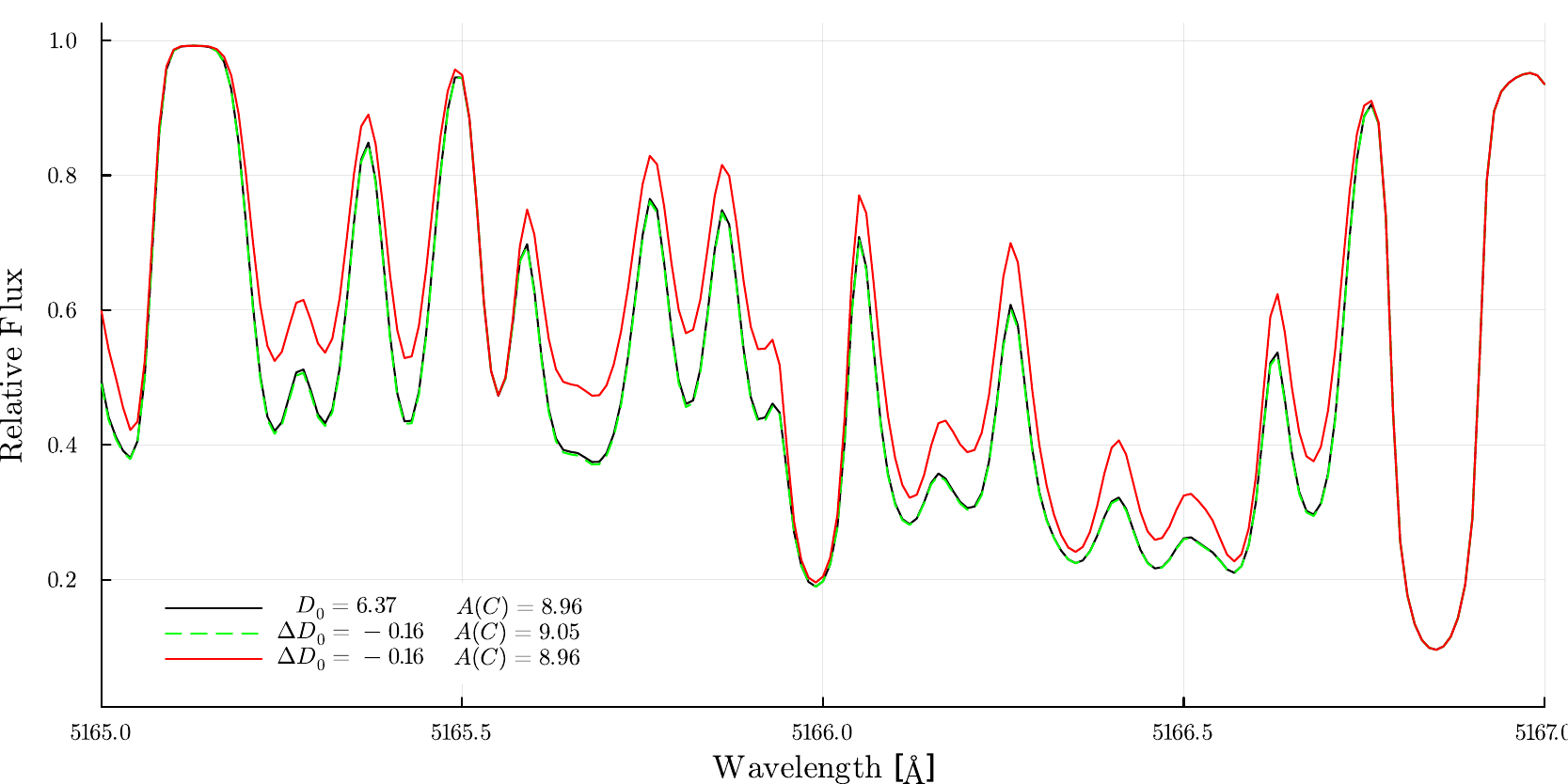}
    \caption{Difference in abundance of carbon in C$_2$ for a star with an effective temperature of 5250~K, log(g) = 2.0, [Fe/H] = 0.5, and [$\alpha$/metals] = 0.1. The black line shows the spectrum synthesised with the fiducial dissociation energy, and the red line with the perturbed dissociation energy. The green line is the spectrum obtained when fitting the perturbed spectrum (red) to the fiducial spectrum (black) by allowing the carbon abundance $A(C)$ to vary.}
    \label{fig: C2 spectra}
\end{figure*}

\noindent
Given the sets of fiducial and perturbed spectra, we found the wavelength with the maximum absolute difference in rectified (continuum-normalised) flux. Although not strictly correct, we assumed that the maximum absolute difference in rectified flux would be a good approximation to where the largest possible abundance differences could be inferred due to an incorrect dissociation energy, giving us a `worst case scenario'. This informed our decision on what wavelength we should focus on when considering how incorrect dissociation energies would translate into abundances. 
We then synthesised fiducial and perturbed spectra for every atmosphere around this `worst case scenario' wavelength ($8\,$\AA\ either side) in order to evaluate the impact on abundances. Note that for MgO and TiO, the flux differences were only calculated for atmospheres with effective temperatures between 4000\,K to 5000\,K. This is the most sensitive regime as these molecules mostly have lines that are relevant at cool temperatures, and very few in hotter atmospheres. 
\\\\
For each atmosphere we then treated the perturbed spectrum as if it were real data, and fit that spectrum using the fiducial dissociation energy, where we allowed the elemental abundances of the constituent species to vary (e.g., C abundance for C$_2$, Mg and O abundances for MgO). Optimisation was performed by minimising the $\chi^2$ difference with the Limited-memory Broyden–Fletcher–Goldfarb–Shannon algorithm and automatic differentiation. The optimisation provided the abundance for the fiducial and perturbed spectra to agree (e.g., identical black and green spectra in Figure \ref{fig: C2 spectra}), allowing us to quantify the abundance difference caused by an incorrect dissociation energy. This process was repeated across approximately 6,200 atmospheres.

\section{Results}
\label{Section: results}

\begin{figure*}

    \includegraphics[width=\textwidth]{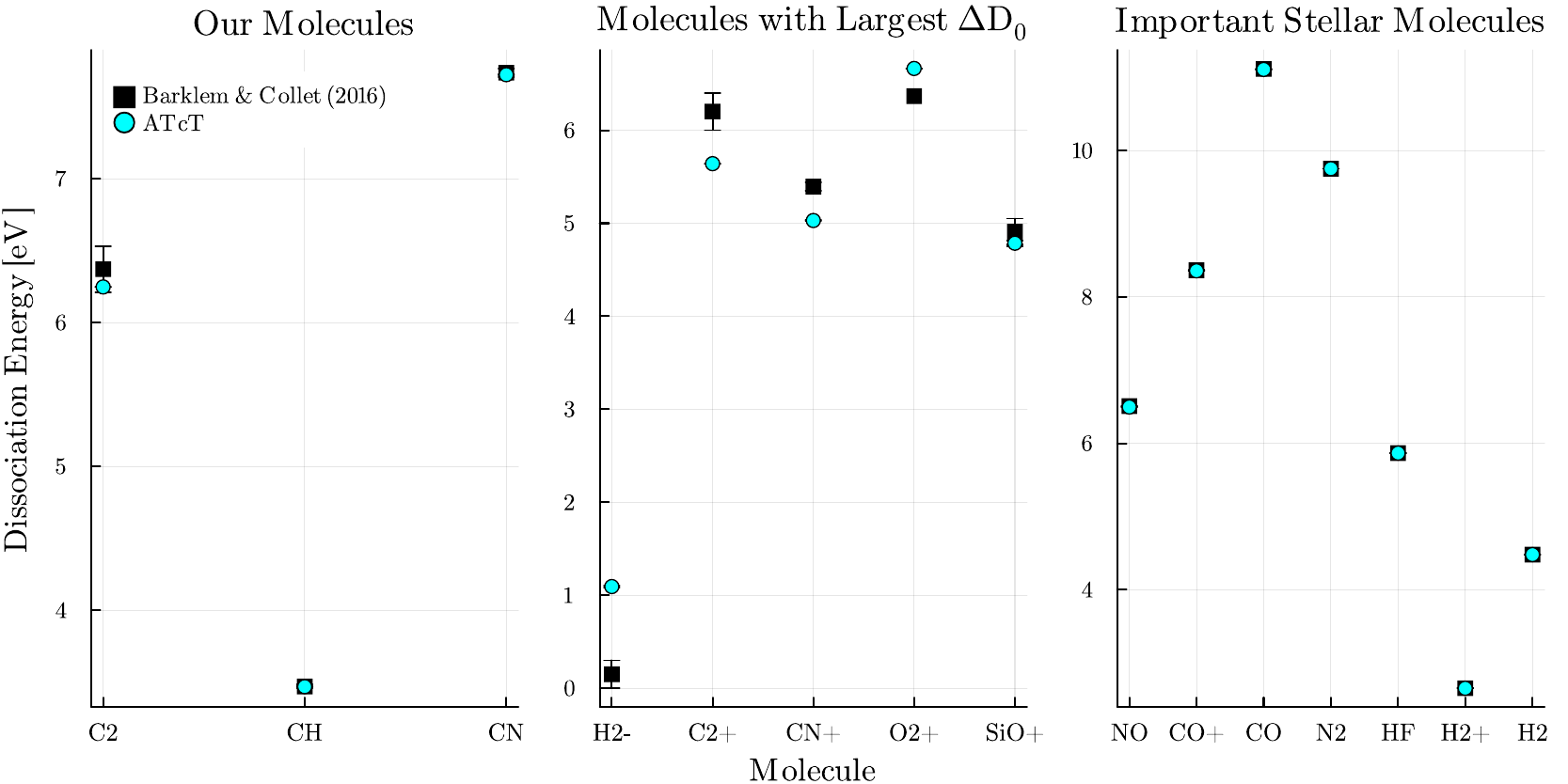}
    \caption{Comparison between the dissociation energies  of \citet{barklem_partition_2016} (black) and ATcT \citep[blue,][]{ruscic_2004}. The left panel includes molecules we investigate in our analysis, the middle panel includes those with the largest difference in dissociation energy, and the right panel includes molecules that have important contributions in stars. ATcT did not have values for MgO and TiO as well as other crucial molecules like FeH.}
    \label{fig: summary_d0}
\end{figure*}

We chose the adopted dissociation energies from \citet{barklem_partition_2016} based on their corresponding uncertainties, not the specific value implemented by various spectral synthesis codes. The more recent collection of dissociation energy values in the Active Thermochemical Tables \citep[ATcT,][]{ruscic_2004} use a thermochemical network to derive reliable values that are self-consistent. A comparison between the dissociation energies of \citet{barklem_partition_2016} and ATcT is shown in Figure \ref{fig: summary_d0} with the first panel displaying the molecules analysed in this paper, the second panel the molecules with the largest differences in dissociation energy and in the third panel some of the molecules that are considered important in stars. The ATcT uncertainties in dissociation energy are at least 1 order of magnitude smaller than those in \citet{barklem_partition_2016} for C$_2$, CN and CH but no dissociation energy values are provided for both TiO and MgO. This is possibly because the compendium was not curated for the purpose of modelling stellar atmospheres or synthesising stellar spectra. Other molecules such as VO and FeH that are essential for stellar models are also not included. However, important stellar molecules present in both ATcT and \citet{barklem_partition_2016} have dissociation energies that differ by less than 0.01~eV despite using independent methodologies. This emphasises the accuracy of the \citet{barklem_partition_2016} compendium although it is overall less precise than ATcT.
\\\\
The largest difference in dissociation energy between \citet{barklem_partition_2016} and ATcT is 0.94~eV for H$_2-$ shown in Figure \ref{fig: summary_d0}. The discrepancy may be from the multiple formation pathways of H$_2-$ that must be observed in experiments. This includes the formation of hydrogen by dissociative attachment and the vibrational excitation of H$_2$. The \citet{herzberg1950molecular} value considered by \citet{barklem_partition_2016} is derived only through observations of dissociative attachments and is an order of magnitude larger than that of \citet{luo2007comprehensive} of 0.15~eV which \citet{barklem_partition_2016} adopted. Both methods did not provide an uncertainty, so the large disagreement between the experimental values was used to derive 0.15~eV as the corresponding uncertainty. 

\begin{figure}[H]

	\includegraphics[width=\columnwidth]{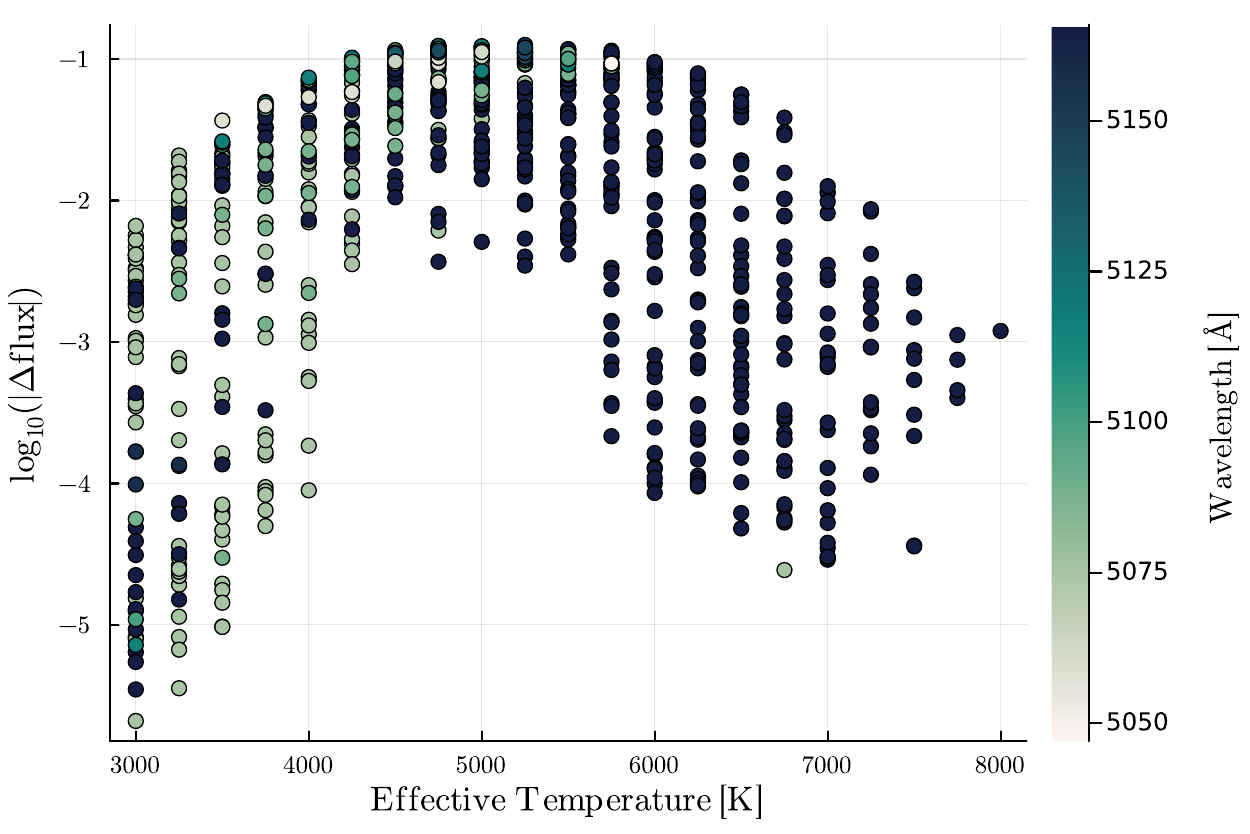}
    \caption{Maximum rectified flux difference between the perturbed spectra and fiducial spectra of each atmosphere sampled for C$_2$. The largest flux difference of 0.13 occurs at 5145\,\AA. Most atmospheres had the largest flux difference at 5166\,\AA.}
    \label{fig: C2 flux difference}
\end{figure}

\begin{figure}[H]

    \centering
    {{\includegraphics[width=\columnwidth]{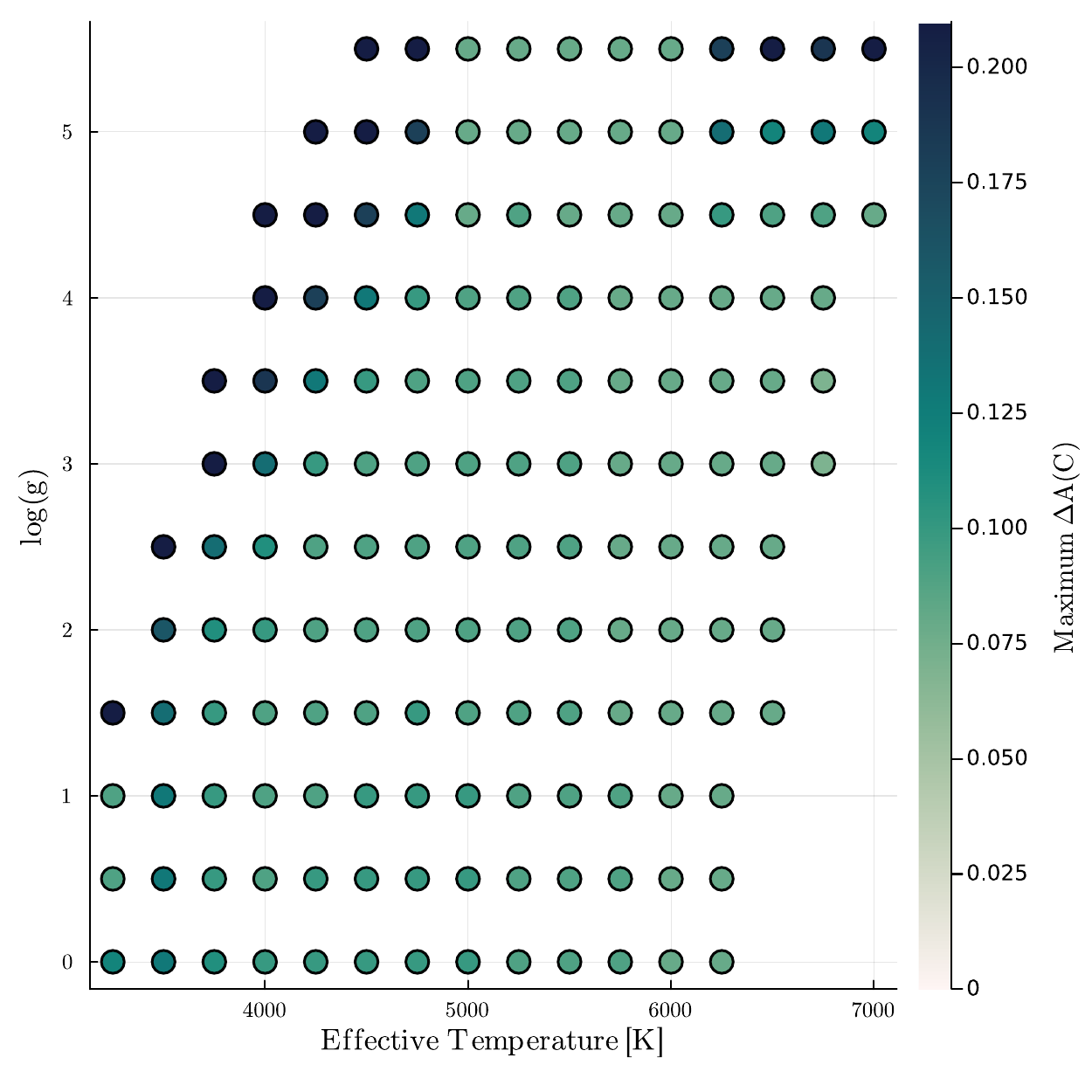} }}%
    \caption{Effective temperature and log(g) across our parameter space. Each point is coloured by the maximum abundance difference. Stars with $\textrm{log}_{10}(|\Delta \textrm{flux}|) < -2.0$  were removed from the figure. Stars with effective temperatures between 4000~K and 5000~K have the maximum abundance difference of 0.20 dex. The minimum abundance difference is 0.0 dex for stars with an effective temperature of 6250~K.}%
    \label{fig: max/min plots}%
    
\end{figure}

\noindent
Despite the significant uncertainties for the dissociation energy of CH, CN, TiO and MgO from \citet{barklem_partition_2016}, all these molecules had flux differences less than 5$\%$. This translates to negligible abundance differences for what is considered the largest difference between the fiducial and perturbed spectra. As a result, the optimisation was only performed on our entire parameter space for C$_2$. However, the flux differences of these four molecules are still significant for specific applications such as detecting exoplanets and simulating the interior of theoretical terrestrial planets \citep{amarsi_carbon_2019}. To accurately determine the mineralogy of the planet, a precision level of $\leq 0.025$ dex is recommended for the stellar
differential abundances of the host star \citep{wang_detailed_2022}. This highlights the need to input the most recent dissociation energies for these four molecules to improve synthetic spectra for wider applications, regardless of the flux difference we have computed. 
\\\\
We found that sometimes the optimiser returned an unrealistic abundance difference when there was almost no change to the spectrum (i.e., less than 10$^{-5}$). In effect, the returned abundance differences represent upper limits on abundance differences instead of measurements. 
For these reasons, in later sections we only consider results where the flux difference is $|\Delta{}\text{normalised~flux}| > 10^{-2}$ in at least one pixel.
\\\\
The uncertainty in the dissociation energy of C$_2$ leads to the largest maximum rectified flux difference. The maximum rectified flux difference and the wavelength at which it occurs is shown for stars with different effective temperatures in Figure \ref{fig: C2 flux difference}. C$_2$ has a maximum rectified flux difference of 0.13 for an atmosphere with an effective temperature of 5250~K, log(g) = 0.0, [Fe/H] = 0.5, [$\alpha$/metals] = 0.0. This occurs at approximately 5145~\AA \hspace{0cm} in Figure \ref{fig: C2 flux difference}. However, most atmospheres showed the largest difference at 5166~\AA. We therefore considered 5166~\AA \hspace{0cm} to be the line within the C$_2$ band (known as the Swan band) most sensitive to changes in the dissociation energy, and used this as our chosen wavelength for the maximum rectified flux \citep{sriramachandran_presence_2016}. This is demonstrated in Figure \ref{fig: C2 spectra} where 5166~\AA\ is the central wavelength. 
\\\\
Inferred carbon abundance differences of up to 0.2\,dex were measured from C$_2$ features (Figure~\ref{fig: max/min plots}), with both a mean and median of 0.08 dex. The maximum abundance difference was observed for two stars: one with an effective temperature of 4250~K, log(g) = 4.5, [Fe/H] = 0.5, and [$\alpha$/metals] = 0.4 and the other with an effective temperature of 4750~K, log(g) = 5.5, [Fe/H] = 0.5, and [$\alpha$/metals] = 0.4. In general, cooler stars ($T_\mathrm{eff} < 5000$\,K) have larger abundance differences than hotter stars ($>$ 6250~K). However, these two stars have a flux difference of 0.012 and 0.013, respectively, an order of magnitude smaller than the maximum rectified flux difference we obtained. The large abundance differences for these stars is likely a result of the slanted continuum leading to the fitted spectrum being slightly offset from the fiducial spectrum.

\section{Discussion}
\label{Section: discussion}

We have evaluated the impact that incorrect dissociation energies for C$_2$, MgO, TiO, CN, and CH have on predicted spectra. These results only relate to elemental abundances as measured from molecular bands, whereas usually elemental abundances are derived from atomic lines wherever possible. However, there are some cases in which carbon can only be measured from molecular features. This includes metal-poor stars, giant stars, or very cool stars, as they have typically have large non-local thermal equilibrium effects for atomic lines and/or heavily blended atomic lines \citep{ryabchikova_using_2022,amarsi_solar_2021}. First we discuss the lesser impact of MgO, TiO, CN, and CH, before we examine the effects of C$_2$.

\subsection{MgO, TiO, CN, and CH}

The maximum flux difference of MgO provides an upper limit for the abundance differences of magnesium and oxygen. However, MgO is a refractory compound, meaning it is resistant to changes at high temperatures and pressures \citep{kloska_gas-phase_2017}. This may explain the smaller flux differences observed between the fiducial and perturbed spectra, despite the change in dissociation energy. The small flux differences may also be attributed to large amounts of blending. The maximum flux difference for MgO occurs in the UV at 3116~\AA\ where there is an overcrowding of lines. This has led to molecular lines with flattened troughs and a relative flux close to 0. 
\\\\
TiO may have small flux differences despite perturbing the dissociation energy, as the abundance of TiO is typically negligible compared to other elements or molecules in stars \citep{hubeny2014theory}. TiO also dissociates above 4000~K possibly leading to lower flux differences for stars with higher effective temperatures \citep{ bidaran_2016}. Despite this, TiO is an important opacity source in cool ($2200-3700$~K) stars because it has many open shells \citep{tennyson2019astronomical}. These open shells provide low-lying electronic states that are thermally occupied, allowing TiO to absorb radiation from stellar interiors over a wide wavelength range. The complexity introduced by these numerous strong transitions in the spectra of cool stars may explain the large spread in flux differences with increasing effective temperature.
\\\\
CN has a maximum flux difference of 0.027 at 4214~\AA\ for an atypical supergiant star ($\log{g} = 0$) at 4750~K with solar abundances. The maximum flux difference corresponds to a carbon abundance difference of 0.04 dex and a nitrogen abundance difference of 0.05 dex. The larger flux differences and abundances for this star may be due to the blending of CN lines. For other stars, the flux differences may be small for CN, as CN and CH are found to have clear anti-correlations in band strengths \citep[except for in metal-poor stars;][]{pancino_low-resolution_2010}. Therefore, as CH has strong molecular bands between 4200~\AA \hspace{0cm} and 4300~\AA, CN has weaker bands at these wavelengths \citep{goswami_ch_2005}. 
\\\\
The maximum flux difference of 0.01 for CH occurs at 3628~\AA\ and translates to a carbon abundance difference of 0.02 dex for a giant metal-poor star ($T_\mathrm{eff} = 4750,\mathrm{K}$, $\log_{10}(g) = 2.5$, $[\mathrm{Fe/H}] = -1.5$). This is expected as most bound-bound transitions for CH appear in the near UV for FGK stars, regardless of their carbon abundance \citep{masseron_ch_2014}. It is promising that the wavelength at which the largest flux difference occurs is not a part of the CH band at 4300~\AA \hspace{0cm}. This band is considered to be one of the most prominent CH bands and holds vital information about a star's magnetic flux concentration \citep{masseron_ch_2014}. CH also has the smallest maximum flux difference compared to the four other molecules we considered. This is likely due to CH having the smallest uncertainty in its dissociation energy in Table \ref{table:dissociation energies}.

\subsection{C$_2$}

The maximum abundance difference between the perturbed and fiducial spectra for C$_2$ across our parameter space is shown in Figure \ref{fig: max/min plots}. In Figure \ref{fig: max/min plots}, the abundance difference reaches a maximum of 0.20 dex between 4000~K and 5000~K. The decrease in abundance difference at temperatures lower than 4000~K may be attributed to the association of other molecules such as CO at lower temperatures, even though C$_2$ does not dissociate \citep{sriramachandran_presence_2016}. CO associates at these lower temperatures since it has a dissociation energy of 11.117 eV, double that of C$_2$ \citep{barklem_partition_2016}. The smaller flux differences and overestimation of abundances above 7000~K are possibly due to C$_2$ being mostly dissociated for both spectra. Stars observed by \citet{Hema_2012} with effective temperatures above 7000~K showed no C$_2$ bands around 5165 \AA.  
\\\\
Atmospheres with effective temperatures from 4500~K to 6000~K provide the most uniform abundance and flux differences across $\log{g}$ values in our parameter space for C$_2$. Stars with an effective temperature below 4500~K and a higher surface gravity (above 2.0) typically had small flux differences. A similar trend was observed for atmospheres with effective temperatures above 6250~K with $\log{g} < 2.5$, which may be because they experience more pressure broadening. Broadened lines would tend to spread flux differences across more pixels, leading to a smaller per-pixel difference. For stars with an effective temperature below 4500~K but with low log(g), the flux differences may have been larger due to weak lines and a slanted continuum. Weak lines are highly sensitive to the continuum placement, which will vary for the fiducial and perturbed spectrum as they have different contributions to the opacity. This leads to larger flux differences and failure by the optimiser to fit the continuum would also result in larger abundance difference. These differences overestimate the impact we would observe due to an incorrect dissociation energy \citep{blanco-cuaresma_modern_2019}. 
\\\\
We found that the abundance differences for C$_2$ increase as a function of metallicity and $[\alpha$/metals]. The increase in abundance difference with $[\alpha$/metals] may be because $\alpha$-enhanced stars have more electron contributions from metals such as Mg and Ca that have low ionisation potentials, increasing the H$^-$ opacity \citep{jofre_gaia_2017}. This decreases the continuum flux and results in a stronger temperature gradient. $\alpha$-enhancement is also more notable for cooler stars, since H- has a low binding energy and is destroyed in hot stellar atmospheres \citep{hubeny2014theory}. This may be why the abundance difference is larger for lower effective temperatures. A similar change in opacity exhibited by $\alpha$-enhancement is observed for metal-rich stars, as they contribute more electrons than metal-poor stars \citep{chisholm_constraining_2019}.  

\section{Conclusions}
\label{sec:conclusions}

We synthesised spectra with both a fiducial and an incorrect (perturbed) dissociation energy to evaluate how incorrect dissociation energies adopted by spectral synthesis codes could translate to systematic biases in inferred elemental abundances. The perturbed dissociation energies were the fiducial values minus their corresponding uncertainty. We found that CH, CN, TiO, and MgO all had a maximum flux difference less than 0.05 dex. This translates to a negligible abundance differences, but could still be important opacity contributors, or for particular science cases (e.g., exoplanet compositions). For C measurements from C$_2$ features, our analysis demonstrated that atmospheres with effective temperatures between 5000~K and 6000~K have C$_2$ carbon abundance differences from 0.05 dex to 0.09 dex (if measured from C$_2$ molecular features near 5166\,\AA). From atomic lines, the \citet{asplund_2020vision_2021} Solar carbon abundance is 8.46 $\pm$ 0.04 dex. In this scenario, the incorrect dissociation energy for C$_2$ translates to a systematic abundance uncertainty twice what is reported for the Sun.
\\\\
If we consider that different dissociation energy values are used in spectral synthesis codes, it becomes difficult to compare theory to observations. For example, \texttt{SME} \citep{SME_1996} uses 6.297 eV, and both \texttt{TurboSpectrum} and \texttt{MOOG} use 6.21 eV for the dissociation energy of C$_2$ \citep{turbospectrum_2012,sneden_2012}. \texttt{TurboSpectrum} and \texttt{MOOG} are just within 1 standard deviation from the 6.371 $\pm$ 0.16 eV adopted by \texttt{Korg}. To the first order we can estimate the code-to-code impact by linear scaling the difference in the dissociation energy to the uncertainty in dissociation energy we adopted. This means that these codes may deviate by up to 60$\%$ from \texttt{Korg} when considering carbon abundances, with smaller discrepancies expected to be observed between \texttt{Korg} and \texttt{SME}.
\\\\
The dissociation energy values collated for the purpose of spectral synthesis are not the most recent values to be calculated, and may be out of date. A more recent C$_2$ dissociation energy of 6.24 $\pm$ 0.02 eV is reported by \citet{visser_new_2019}. They used a combination of mixing experiments and multi-reference configuration interaction calculations to derive this value. Their dissociation energy is in agreement with spectroscopic results from the NEAT dataset \citep{csaszar_network_2010} and Active Thermochemical Tables \citep{ruscic_2004} as well as theoretical results from CCSD(T) \citep{karton_atomization_2009,feller_improved_2013}. It is within 1 standard deviation of that recommended by \citet{luo2007comprehensive}. The uncertainty in the dissociation energy determined by \citet{visser_new_2019} also considers relativistic effects and contributions from high order excitation states. This has contributed to a precise dissociation energy value for C$_2$, leading to a negligible abundance uncertainty. The C$_2$ dissociation energy from \citet{visser_new_2019} is likely more accurate and precise than the values recommended by \citet{luo2007comprehensive}. 
\\\\
These are significant abundance differences between the spectra of differing dissociation energies for C$_2$. The significant impact of incorrect dissociation energies on Solar-like stars highlights the importance for a consensus on the dissociation energy amongst spectral synthesis codes. Dissociation energy values adopted by these codes also need to be updated so we can be confident that the spectra produced are reliable and all inferences are accurate. Our analysis could also be extended by fitting synthetic spectra around an atomic carbon line and a molecular C$_2$ line for different stars. This would allow us to optimise the carbon abundance as well as the dissociation energy simultaneously. We could improve dissociation energy measurements using synthetic spectra instead of experimental techniques.

\section*{Acknowledgements}
We thank Paul Barklem for his enthusiasm for the project and invaluable feedback on drafts of this paper. We also thank Chris Sneden for his advice. 
\\\\
This work has made use of the VALD database, operated at Uppsala University, the Institute of Astronomy RAS in Moscow, and the University of Vienna.

\section*{Data Availability}
All input data used are publicly available or can be requested from the lead author.

\bibliographystyle{mnras}
\bibliography{mnemonic,main} 

\begin{thebibliography}{}
\makeatletter
\relax
\def\mn@urlcharsother{\let\do\@makeother \do\$\do\&\do\#\do\^\do\_\do\%\do\~}
\def\mn@doi{\begingroup\mn@urlcharsother \@ifnextchar [ {\mn@doi@} {\mn@doi@[]}}
\def\mn@doi@[#1]#2{\def\@tempa{#1}\ifx\@tempa\@empty \href {http://dx.doi.org/#2} {doi:#2}\else \href {http://dx.doi.org/#2} {#1}\fi \endgroup}
\def\mn@eprint#1#2{\mn@eprint@#1:#2::\@nil}
\def\mn@eprint@arXiv#1{\href {http://arxiv.org/abs/#1} {{\tt arXiv:#1}}}
\def\mn@eprint@dblp#1{\href {http://dblp.uni-trier.de/rec/bibtex/#1.xml} {dblp:#1}}
\def\mn@eprint@#1:#2:#3:#4\@nil{\def\@tempa {#1}\def\@tempb {#2}\def\@tempc {#3}\ifx \@tempc \@empty \let \@tempc \@tempb \let \@tempb \@tempa \fi \ifx \@tempb \@empty \def\@tempb {arXiv}\fi \@ifundefined {mn@eprint@\@tempb}{\@tempb:\@tempc}{\expandafter \expandafter \csname mn@eprint@\@tempb\endcsname \expandafter{\@tempc}}}

\bibitem[\protect\citeauthoryear{Amarsi, Lind, Asplund, Barklem  \& Collet}{Amarsi et~al.}{2016}]{amarsi_non-lte_2016}
Amarsi A.~M.,  Lind K.,  Asplund M.,  Barklem P.~S.,   Collet R.,  2016, \mn@doi [Monthly Notices of the Royal Astronomical Society] {10.1093/mnras/stw2077}, 463, 1518

\bibitem[\protect\citeauthoryear{Amarsi, Nissen  \& Skúladóttir}{Amarsi et~al.}{2019}]{amarsi_carbon_2019}
Amarsi A.~M.,  Nissen P.~E.,   Skúladóttir A.,  2019, \mn@doi [Astronomy \& Astrophysics] {10.1051/0004-6361/201936265}, 630, 1

\bibitem[\protect\citeauthoryear{Amarsi, Grevesse, Asplund  \& Collet}{Amarsi et~al.}{2021}]{amarsi_solar_2021}
Amarsi A.~M.,  Grevesse N.,  Asplund M.,   Collet R.,  2021, \mn@doi [Astronomy \& Astrophysics] {10.1051/0004-6361/202141384}, 656, A113

\bibitem[\protect\citeauthoryear{Arentsen, Placco, Lee, Aguado, Martin, Starkenburg  \& Yoon}{Arentsen et~al.}{2022}]{arentsen_inconsistency_2022}
Arentsen A.,  Placco V.~M.,  Lee Y.~S.,  Aguado D.~S.,  Martin N.~F.,  Starkenburg E.,   Yoon J.,  2022, \mn@doi [Monthly Notices of the Royal Astronomical Society] {10.1093/mnras/stac2062}, 515, 4082

\bibitem[\protect\citeauthoryear{{Asplund}, {Amarsi}  \& {Grevesse}}{{Asplund} et~al.}{2021}]{asplund_2020vision_2021}
{Asplund} M.,  {Amarsi} A.~M.,   {Grevesse} N.,  2021, \mn@doi [\aap] {10.1051/0004-6361/202140445}, \href {https://ui.adsabs.harvard.edu/abs/2021A&A...653A.141A} {653, A141}

\bibitem[\protect\citeauthoryear{Barklem \& Collet}{Barklem \& Collet}{2016}]{barklem_partition_2016}
Barklem P.~S.,  Collet R.,  2016, \mn@doi [Astronomy \& Astrophysics] {10.1051/0004-6361/201526961}, 588, 1

\bibitem[\protect\citeauthoryear{Bartlett \& Silver}{Bartlett \& Silver}{1974}]{bartlett_1974}
Bartlett R.~J.,  Silver D.~M.,  1974, \mn@doi [International Journal of Quantum Chemistry] {https://doi.org/10.1002/qua.560080831}, 8, 271

\bibitem[\protect\citeauthoryear{Beers \& Christlieb}{Beers \& Christlieb}{2005}]{beers_discovery_2005}
Beers T.~C.,  Christlieb N.,  2005, \mn@doi [Annual Review of Astronomy and Astrophysics] {10.1146/annurev.astro.42.053102.134057}, 43, 531

\bibitem[\protect\citeauthoryear{Bidaran, Mirtorabi  \& Azizi}{Bidaran et~al.}{2016}]{bidaran_2016}
Bidaran B.,  Mirtorabi M.~T.,   Azizi F.,  2016, \mn@doi [Monthly Notices of the Royal Astronomical Society] {10.1093/mnras/stw051}, 457, 2043

\bibitem[\protect\citeauthoryear{Blanco-Cuaresma}{Blanco-Cuaresma}{2019}]{blanco-cuaresma_modern_2019}
Blanco-Cuaresma S.,  2019, \mn@doi [Monthly Notices of the Royal Astronomical Society] {10.1093/mnras/stz549}, 486, 2075

\bibitem[\protect\citeauthoryear{Chisholm, Rigby, Bayliss, Berg, Dahle, Gladders  \& Sharon}{Chisholm et~al.}{2019}]{chisholm_constraining_2019}
Chisholm J.,  Rigby J.~R.,  Bayliss M.,  Berg D.~A.,  Dahle H.,  Gladders M.,   Sharon K.,  2019, \mn@doi [The Astrophysical Journal] {10.3847/1538-4357/ab3104}, 882, 182

\bibitem[\protect\citeauthoryear{Costes \& Naulin}{Costes \& Naulin}{1994}]{costes_naulin_1994}
Costes M.,  Naulin C.,  1994, \mn@doi [International Astronomical Union Colloquium] {10.1017/S0252921100021370}, 146, 250–264

\bibitem[\protect\citeauthoryear{Császár \& Furtenbacher}{Császár \& Furtenbacher}{2010}]{csaszar_network_2010}
Császár A.~G.,  Furtenbacher T.,  2010, \mn@doi [Chemistry (Weinheim an Der Bergstrasse, Germany)] {10.1002/chem.200903252}, 16, 4826

\bibitem[\protect\citeauthoryear{Curtiss, Raghavachari, Trucks  \& Pople}{Curtiss et~al.}{1991}]{curtiss_gaussian2_1991}
Curtiss L.~A.,  Raghavachari K.,  Trucks G.~W.,   Pople J.~A.,  1991, \mn@doi [The Journal of Chemical Physics] {10.1063/1.460205}, 94, 7221

\bibitem[\protect\citeauthoryear{Curtiss, Redfern  \& Raghavachari}{Curtiss et~al.}{2007}]{curtiss_gaussian-4_2007}
Curtiss L.~A.,  Redfern P.~C.,   Raghavachari K.,  2007, \mn@doi [The Journal of Chemical Physics] {10.1063/1.2436888}, 126, 084108

\bibitem[\protect\citeauthoryear{Feller, Peterson  \& Ruscic}{Feller et~al.}{2013}]{feller_improved_2013}
Feller D.,  Peterson K.~A.,   Ruscic B.,  2013, \mn@doi [Theoretical Chemistry Accounts] {10.1007/s00214-013-1407-z}, 133, 1407

\bibitem[\protect\citeauthoryear{Goswami}{Goswami}{2005}]{goswami_ch_2005}
Goswami A.,  2005, \mn@doi [Monthly Notices of the Royal Astronomical Society] {10.1111/j.1365-2966.2005.08917.x}, 359, 531

\bibitem[\protect\citeauthoryear{Gustafsson, Edvardsson, Eriksson, Jorgensen, Nordlund  \& Plez}{Gustafsson et~al.}{2008}]{gustafsson_grid_2008}
Gustafsson B.,  Edvardsson B.,  Eriksson K.,  Jorgensen U.~G.,  Nordlund A.,   Plez B.,  2008, \mn@doi [Astronomy \& Astrophysics] {10.1051/0004-6361:200809724}, 486, 951

\bibitem[\protect\citeauthoryear{Hema, Pandey  \& Lambert}{Hema et~al.}{2012}]{Hema_2012}
Hema B.~P.,  Pandey G.,   Lambert D.~L.,  2012, \mn@doi [The Astrophysical Journal] {10.1088/0004-637X/747/2/102}, 747, 102

\bibitem[\protect\citeauthoryear{Herzberg}{Herzberg}{1950}]{herzberg1950molecular}
Herzberg G.,  1950, Molecular Spectra and Molecular Structure - Vol I.
No.~v. 1, Read Books Limited, \url {https://books.google.com.au/books?id=W2Z8CgAAQBAJ}

\bibitem[\protect\citeauthoryear{{Hourihane} et~al.,}{{Hourihane} et~al.}{2023}]{ges}
{Hourihane} A.,  et~al., 2023, \mn@doi [\aap] {10.1051/0004-6361/202345910}, \href {https://ui.adsabs.harvard.edu/abs/2023A&A...676A.129H} {676, A129}

\bibitem[\protect\citeauthoryear{Hubeny \& Mihalas}{Hubeny \& Mihalas}{2014}]{hubeny2014theory}
Hubeny I.,  Mihalas D.,  2014, Theory of Stellar Atmospheres: An Introduction to Astrophysical Non-equilibrium Quantitative Spectroscopic Analysis.
Princeton Series in Astrophysics, Princeton University Press, \url {https://books.google.com.au/books?id=TmuYDwAAQBAJ}

\bibitem[\protect\citeauthoryear{Jofre et~al.,}{Jofre et~al.}{2017}]{jofre_gaia_2017}
Jofre P.,  et~al., 2017, \mn@doi [Astronomy \& Astrophysics] {10.1051/0004-6361/201629833}, 601, A38

\bibitem[\protect\citeauthoryear{Jofré et~al.,}{Jofré et~al.}{2015}]{jofre_gaia_2015}
Jofré P.,  et~al., 2015, \mn@doi [Astronomy \& Astrophysics] {10.1051/0004-6361/201526604}, 582, A81

\bibitem[\protect\citeauthoryear{Karton, Tarnopolsky  \& Martin}{Karton et~al.}{2009}]{karton_atomization_2009}
Karton A.,  Tarnopolsky A.,   Martin J. M.~L.,  2009, \mn@doi [Molecular Physics] {10.1080/00268970802708959}, 107, 977

\bibitem[\protect\citeauthoryear{Kloska \& Fortenberry}{Kloska \& Fortenberry}{2017}]{kloska_gas-phase_2017}
Kloska K.,  Fortenberry R.,  2017, \mn@doi [Monthly Notices of the Royal Astronomical Society] {10.1093/mnras/stx2912}, 474

\bibitem[\protect\citeauthoryear{{Kupka}, {Piskunov}, {Ryabchikova}, {Stempels}  \& {Weiss}}{{Kupka} et~al.}{1999}]{kupka_1999}
{Kupka} F.,  {Piskunov} N.,  {Ryabchikova} T.~A.,  {Stempels} H.~C.,   {Weiss} W.~W.,  1999, \mn@doi [\aaps] {10.1051/aas:1999267}, \href {https://ui.adsabs.harvard.edu/abs/1999A&AS..138..119K} {138, 119}

\bibitem[\protect\citeauthoryear{Luo}{Luo}{2007}]{luo2007comprehensive}
Luo Y.,  2007, Comprehensive Handbook of Chemical Bond Energies.
CRC Press, \url {https://books.google.com.au/books?id=kM3mwD4y\_TAC}

\bibitem[\protect\citeauthoryear{Masseron et~al.,}{Masseron et~al.}{2014}]{masseron_ch_2014}
Masseron T.,  et~al., 2014, \mn@doi [Astronomy \& Astrophysics] {10.1051/0004-6361/201423956}, 571, A47

\bibitem[\protect\citeauthoryear{Pancino, Rejkuba, Zoccali  \& Carrera}{Pancino et~al.}{2010}]{pancino_low-resolution_2010}
Pancino E.,  Rejkuba M.,  Zoccali M.,   Carrera R.,  2010, \mn@doi [Astronomy \& Astrophysics] {10.1051/0004-6361/201014383}, 524, A44

\bibitem[\protect\citeauthoryear{{Plez}}{{Plez}}{2012}]{turbospectrum_2012}
{Plez} B.,  2012, {Turbospectrum: Code for spectral synthesis}, Astrophysics Source Code Library, record ascl:1205.004 (\mn@eprint {ascl} {1205.004})

\bibitem[\protect\citeauthoryear{Pople, Head‐Gordon, Fox, Raghavachari  \& Curtiss}{Pople et~al.}{1989}]{pople_gaussian1_1989}
Pople J.~A.,  Head‐Gordon M.,  Fox D.~J.,  Raghavachari K.,   Curtiss L.~A.,  1989, \mn@doi [The Journal of Chemical Physics] {10.1063/1.456415}, 90, 5622

\bibitem[\protect\citeauthoryear{Raghavachari}{Raghavachari}{1985}]{raghavachari_augmented_1985}
Raghavachari K.,  1985, \mn@doi [The Journal of Chemical Physics] {10.1063/1.448718}, 82, 4607

\bibitem[\protect\citeauthoryear{Ruscic et~al.,}{Ruscic et~al.}{2004}]{ruscic_2004}
Ruscic B.,  et~al., 2004, \mn@doi [The Journal of Physical Chemistry A] {10.1021/jp047912y}, 108, 9979

\bibitem[\protect\citeauthoryear{Ruscic, Feller  \& Peterson}{Ruscic et~al.}{2014}]{ruscic_active_2014}
Ruscic B.,  Feller D.,   Peterson K.~A.,  2014, \mn@doi [Theoretical Chemistry Accounts] {10.1007/s00214-013-1415-z}, 133, 1415

\bibitem[\protect\citeauthoryear{Ryabchikova, Piskunov  \& Pakhomov}{Ryabchikova et~al.}{2022}]{ryabchikova_using_2022}
Ryabchikova T.,  Piskunov N.,   Pakhomov Y.,  2022, \mn@doi [Atoms] {10.3390/atoms10040103}, 10, 103

\bibitem[\protect\citeauthoryear{{Sneden}, {Bean}, {Ivans}, {Lucatello}  \& {Sobeck}}{{Sneden} et~al.}{2012}]{sneden_2012}
{Sneden} C.,  {Bean} J.,  {Ivans} I.,  {Lucatello} S.,   {Sobeck} J.,  2012, {MOOG: LTE line analysis and spectrum synthesis}, Astrophysics Source Code Library, record ascl:1202.009 (\mn@eprint {ascl} {1202.009})

\bibitem[\protect\citeauthoryear{Sriramachandran, Sindhan, Ramaswamy  \& Shanmugavel}{Sriramachandran et~al.}{2016}]{sriramachandran_presence_2016}
Sriramachandran P.,  Sindhan R.,  Ramaswamy S.,   Shanmugavel R.,  2016, \mn@doi [New Astronomy] {10.1016/j.newast.2016.03.004}, 48, 16

\bibitem[\protect\citeauthoryear{Tennyson}{Tennyson}{2019}]{tennyson2019astronomical}
Tennyson J.,  2019, Astronomical Spectroscopy: An Introduction To The Atomic And Molecular Physics Of Astronomical Spectroscopy (Third Edition).
Advanced Textbooks In Physics, World Scientific Publishing Company, \url {https://books.google.com.au/books?id=neaWDwAAQBAJ}

\bibitem[\protect\citeauthoryear{{Valenti} \& {Piskunov}}{{Valenti} \& {Piskunov}}{1996}]{SME_1996}
{Valenti} J.~A.,  {Piskunov} N.,  1996, \aaps, \href {https://ui.adsabs.harvard.edu/abs/1996A&AS..118..595V} {118, 595}

\bibitem[\protect\citeauthoryear{Visser, Beck, Bornhauser, Knopp, Van~Bokhoven, Radi, Gourlaouen  \& Marquardt}{Visser et~al.}{2019}]{visser_new_2019}
Visser B.,  Beck M.,  Bornhauser P.,  Knopp G.,  Van~Bokhoven J.~A.,  Radi P.,  Gourlaouen C.,   Marquardt R.,  2019, \mn@doi [Molecular Physics] {10.1080/00268976.2018.1564849}, 117, 1645

\bibitem[\protect\citeauthoryear{Wang, Quanz, Yong, Liu, Seidler, Acuña  \& Mojzsis}{Wang et~al.}{2022}]{wang_detailed_2022}
Wang H.~S.,  Quanz S.~P.,  Yong D.,  Liu F.,  Seidler F.,  Acuña L.,   Mojzsis S.~J.,  2022, \mn@doi [Monthly Notices of the Royal Astronomical Society] {10.1093/mnras/stac1119}, 513, 5829

\bibitem[\protect\citeauthoryear{WeiGuo, QunChao  \& WeiYi}{WeiGuo et~al.}{2007}]{weiguo_accurate_2007}
WeiGuo S.,  QunChao F.,   WeiYi R.,  2007, \mn@doi [Science in China Series G: Physics, Mechanics and Astronomy volume] {10.1007/s11433-007-0065-3}, 50, 611

\bibitem[\protect\citeauthoryear{{Wheeler}, {Abruzzo}, {Casey}  \& {Ness}}{{Wheeler} et~al.}{2022}]{korg_2022}
{Wheeler} A.~J.,  {Abruzzo} M.~W.,  {Casey} A.~R.,   {Ness} M.~K.,  2022, {Korg: 1D local thermodynamic equilibrium stellar spectral synthesis}, Astrophysics Source Code Library, record ascl:2211.016 (\mn@eprint {ascl} {2211.016})

\bibitem[\protect\citeauthoryear{Wilkinson}{Wilkinson}{1963}]{wilkinson_1963}
Wilkinson P.,  1963, The Astrophysical Journal, 138, 778

\makeatother
\end{thebibliography}

\bsp	
\label{lastpage}
\end{document}